\documentclass[12pt,a4paper]{article}
\usepackage[utf8]{inputenc}
\usepackage{amsmath}
\usepackage{amsfonts}
\usepackage{amssymb}
\usepackage{graphicx, datetime, abstract}
\usepackage{float}
\usepackage{fullpage}
\usepackage{caption}
\usepackage{subcaption}
\usepackage{todonotes}
\usepackage{setspace}
\usepackage{resizegather}
\usepackage{multirow, hhline}
\usepackage{color,soul}
\pdfoutput=1

\begin{document}
	\title{
			\begin{flushright}\ \vskip -2cm {\small{\em DCPT-16/43}}\end{flushright}
		Salty popcorn in a homogeneous low-dimensional toy model of holographic QCD}
	\author{Matthew Elliot-Ripley\\[10pt]
		{\em \normalsize Department of Mathematical Sciences, }\\{\em \normalsize Durham University, South Rd, Durham, UK}\\[10pt]
		{\normalsize m.k.i.d.elliot-ripley@durham.ac.uk}}
		\date{October 2016}
	\maketitle
	\vspace{15pt}
	\begin{abstract}
	Recently, a homogeneous ansatz has been used to study cold dense nuclear matter in the Sakai-Sugimoto model of holographic QCD. To justify this homogeneous approximation we here investigate a homogeneous ansatz within a low-dimensional toy version of Sakai-Sugimoto to study finite baryon density configurations and compare it to full numerical solutions. We find the ansatz corresponds to enforcing a dyon salt arrangement in which the soliton solutions are split into half-soliton layers. Within this ansatz we find analogues of the proposed baryonic popcorn transitions, in which solutions split into multiple layers in the holographic direction. The homogeneous results are found to qualitatively match the full numerical solutions, lending confidence to the homogeneous approximations of the full Sakai-Sugimoto model. In addition, we find exact compact solutions in the high density, flat space limit which demonstrate the existence of further popcorn transitions to three layers and beyond.
	\end{abstract}
	
	\newpage 

\clearpage
\graphicspath{{../THESIS/pics/Ch4/}}
\section{Introduction}
The Sakai-Sugimoto model \cite{Sakai:2004cn,Sakai:2005yt} is the leading top-down model for holographic QCD. The model can be formulated in the low-energy limit as a Yang-Mills-Chern-Simons theory in $(4+1)$-dimensions in a warped AdS-like spacetime, with bulk instantons corresponding to Skyrmions (i.e. baryons) on the boundary. Investigation of the soliton solutions in this model is difficult both analytically and numerically, and to date only the single soliton has been computed \cite{Bolognesi:2013nja}.

The investigation of finite-density solutions in the Sakai-Sugimoto model is crucial for understanding cold, dense QCD in the holographic setting. In the limit of a large number of colours it is expected for cold nuclear matter to form a crystalline structure, and this should be captured in the holographic description. In addition, the way in which bulk solitons explore the holographic direction as baryon density increases has been proposed as a method of realising the quarkyonic phase \cite{McLerran:2007qj} (in which there is a quark Fermi sea with baryonic Fermi surface) and chiral symmetry restoration \cite{deBoer:2012ij}.

The two main proposals for how the holographic direction is explored at high densities are the dyon salt \cite{Rho:2009ym} and baryonic popcorn \cite{Kaplunovsky:2012gb, Kaplunovsky:2015zsa}. In the dyon salt phase the soliton solutions split into half-soliton constituents analogous to the splitting of flat space Yang-Mills calorons (periodic instantons) into oppositely charged dyons. On the other hand, baryonic popcorn involves a series of transitions occurring in which the three-dimensional soliton crystal develops extra layers in the holographic direction.

Recently, low-dimensional toy models for the Sakai-Sugimoto model have been investigated to study these phenomena. In \cite{Bolognesi:2013jba} a baby Skyrme model in a warped spacetime was used as a low-dimensional analogue, and configurations corresponding to both the dyon salt and baryonic popcorn were found numerically. A series of critical densities were found at which popcorn transitions occur, resulting in more energetically favourable solutions when compared to the dyon salt configurations. A similar model was studied in \cite{Elliot-Ripley:2015cma} where the baby Skyrme term was replaced with a vector meson term (arguably a better low-dimensional analogue of the Chern-Simons term) and similar results were obtained.

In addition to low-dimensional toy models, homogeneous ansatze have recently been used as a way of shedding light on high density solutions of the five-dimensional Sakai-Sugimoto model. In \cite{Elliot-Ripley:2016uwb} a homogeneous ansatz for the five-dimensional Sakai-Sugimoto model was developed which managed to successfully reproduce analogues of baryonic popcorn, dyon salt and soliton bag configurations. The homogeneous ansatz reduced the theory down to an effective kink model in one dimension, and predicted a popcorn transition at a critical baryon number density that is well below the density required to form a dyon salt. No further pops were found beyond the first transition, and this was explained by the formation of a soliton bag whose interior is filled with abelian electric potential, with the instanton number density localised around the surface of the bag. The preference of two layered solutions over solutions with more layers was also found in \cite{Preis:2016fsp} in which a different homogeneous approximation was applied.

These results differ qualitatively from the numerical study of a low-dimensional toy models for the Sakai-Sugimoto model in which multiple popcorn transitions were predicted and observed.
This could be for a variety of reasons, including the different dimensionality of the models, or the presence of a nontrivial gauge potential in the higher dimensional model that is absent in the toy model. To rule out the homogeneous approximation as a cause for the different qualitative behaviour,
we here return to the low-dimensional toy model studied in \cite{Bolognesi:2013jba} and search for a homogeneous ansatz to capture the qualitative features obtained in the full numerical study. We will find that, unlike in the Sakai-Sugimoto model, we can find an ansatz of spatially dependent fields that give rise to energy and baryon number densities that are homogeneous in the non-holographic direction. We find an ansatz that can only give solutions in the dyon salt phase, and find baryonic popcorn transitions within this sector, which we refer to as ``salty popcorn''. We analyse the behaviour of the model in the limiting cases of high baryon number density and flat space, and find counter-intuitive scaling results. We also find that the homogeneous ansatz provides a good approximation to true solutions in the low-dimensional model by comparing the homogeneous solutions to full numerical calculations.
This is used as evidence to justify the use of a recent homogeneous approximation within the five-dimensional Sakai-Sugimoto model \cite{Elliot-Ripley:2016uwb}.

\section{A homogeneous baby Skyrme model}
The low-dimensional toy model of holographic QCD studied in \cite{Bolognesi:2013jba} is an $O(3)$-sigma model in a warped spacetime, stabilised by a baby Skyrme term. The spacetime metric is given by
\begin{equation}\label{metricch3}
ds^2 = H(z)(-dt ^2+dx^2) + \frac{1}{H(z)}dz^2\,,
\end{equation}
where $H(z)= \left(1+z^2\right)^{1/2}$ is the spacetime warp factor, and the theory is defined by the action
\begin{equation}\label{bsactionch1}
S = \int \left( -\frac{1}{2}\partial_\mu\pmb{\phi}\cdot\partial^\mu\pmb{\phi} - \frac{\kappa^2}{4}(\partial_\mu\pmb{\phi}\times\partial_\nu\pmb{\phi})\cdot(\partial^\mu\pmb{\phi}\times\partial^\nu\pmb{\phi}) \right)\sqrt{H}\, dx\, dz\, dt
\end{equation}
where the first term is that of the $O(3)$-sigma model, and the second term is the baby Skyrme term with constant coefficient $\kappa^2$. Greek indices run over the bulk spactime coordinates $t$, $x$ and $z$.  The field $\pmb{\phi} = (\phi_1 , \phi_2 , \phi_3)$ is a three component unit vector. We will refer to $\pmb{\phi}$ as the pion field since the baby Skyrme model can be seen as a 2-dimensional analogue of the Skyrme model where components of the corresponding $\pmb{\phi}$ field play the role of pions. The associated static energy of this model is
\begin{equation}\label{BSEch1}
E = \frac{1}{2}\int\,\left( \frac{1}{\sqrt{H}}|\partial_x\pmb{\phi}|^2 + H^{3/2} |\partial_z\pmb{\phi}|^2 + \kappa^2\sqrt{H} |\partial_x\pmb{\phi}\times\partial_z\pmb{\phi}|^2 \right)\, dx\, dz\, .
\end{equation}

For finite energy we then require $\pmb{\phi} \rightarrow (0,0,1)$ as $x^2+z^2\rightarrow\infty$. This allows us to compactify space from $\mathbb{R}^2$ to $S^2$, meaning the pion field is now a map $\pmb{\phi}:S^2\rightarrow S^2$ with an associated winding number and topological charge
\begin{equation}
B = \int\, B^0 \, \sqrt{H}\, dx\, dz = -\frac{1}{4\pi}\int\,\pmb{\phi}\cdot(\partial_x\pmb{\phi}\times\partial_z\pmb{\phi})\, dx\, dz\,,
\end{equation}
which defines the instanton number of the planar $O(3)$-sigma model. This topological charge we identify with the baryon number of the configuration.

Our homogeneous ansatz for this model will be to write:
\begin{equation}
\pmb{\phi}(x,z) = 
\begin{cases}
\left(\sin\!{f(z)}\cos\!{(\pi\rho x)}, -\sin\!{f(z)}\sin\!{(\pi\rho x)}, \cos\!{f(z)}\right)\,,  &z\ge 0\\
\left(\sin\!{f(z)}\cos\!{(\pi\rho x)}, +\sin\!{f(z)}\sin\!{(\pi\rho x)}, \cos\!{f(z)}\right)\,, &z\le 0\\
\end{cases}
\end{equation}
where $f(z)$ is some profile function dependent on the holographic coordinate and $\rho$ is the baryon number density in the $x$-direction. The sign flip in the second component over $z=0$ is to ensure the configuration has non-zero topological charge, and is related to the internal phases of the constituent solitons in the non-homogeneous model. The sign flip can also be seen as the analogue of the change in sign of $F_{ij}$ in the homogeneous kink model of \cite{Elliot-Ripley:2016uwb}. The corresponding energy and soliton number (per unit length in the $x$-direction) are then
\begin{align}
E &= \int_0^\infty\,\left( \frac{\pi^2\rho^2}{H}\sin^2\!{f} + H\left(f^\prime\right)^2 + \kappa^2\pi^2\rho^2\left(f^\prime\right)^2\sin^2\!{f}\right)\sqrt{H}\, dz\,, \label{Ehomog1}\\
\rho &= -\frac{\rho}{2}\int_0^\infty\,f^\prime\sin\!{f}\, dz \label{Bhomog1}\,.
\end{align}
We impose boundary conditions $f(0)=\pi$ and $f(\pm\infty)=0$, which then imply the pion field satisfies $\pmb{\phi}(x,0) = (0,0,-1)$ and $\pmb{\phi}(x,\pm\infty) = (0,0,1)$. Since these field values in the baby Skyrme model correspond to the centre of a soliton and the vacuum respectively, we interpret our ansatz as a homogeneous line of soliton material located at $z=0$. We interpret the soliton number per unit length as the baryon number density.

We should note that this ansatz corresponds to the dyonic salt configuration, which is an analogue of the splitting of high-density calorons into monopole constituents with half the topological charge. Here, the baryon number density is proportional to $\sin\!{f}$, and so vanishes at $z=0$ and $z=\pm\infty$. The baryon number density must therefore be localised around two points away from the origin, rather than be localised around $z=0$, which is characteristic of the splitting behaviour of the dyonic salt. This already leads to problems with the homogeneous ansatz, in that we cannot use it to predict whether a baryonic popcorn phase is preferred over a dyon salt phase at high densities.

This presents another significant difference between the homogeneous approximations of the low-dimensional toy model and the five-dimensional Sakai-Sugimoto model. In the latter, a zero-curvature condition on the five-dimensional gauge fields was used to approximate solutions by a fully spatially homogeneous kink field. This was possible due to the boundary conditions on the gauge field at $z=\pm\infty$. In the toy model, however, we do not have a gauge theory and there is no analogous zero-curvature condition. In fact, it can be shown that any field ansatz in the toy model periodic in $x$ with boundary conditions $\pmb{\phi}(x,0)=(0,0,-1)$ and $\pmb{\phi}(x,\infty)=(0,0,1)$ must correspond to a salty ansatz (see Appendix~\ref{ap:salt}).

\begin{figure}[p]
	\centering
	\includegraphics[width=0.48\textwidth]{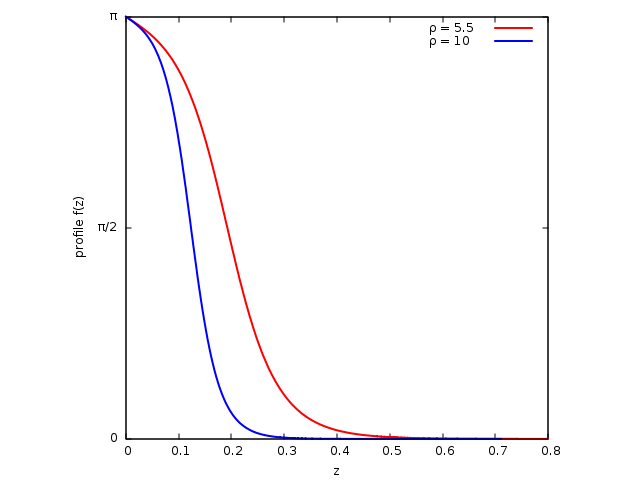}
	\includegraphics[width=0.48\textwidth]{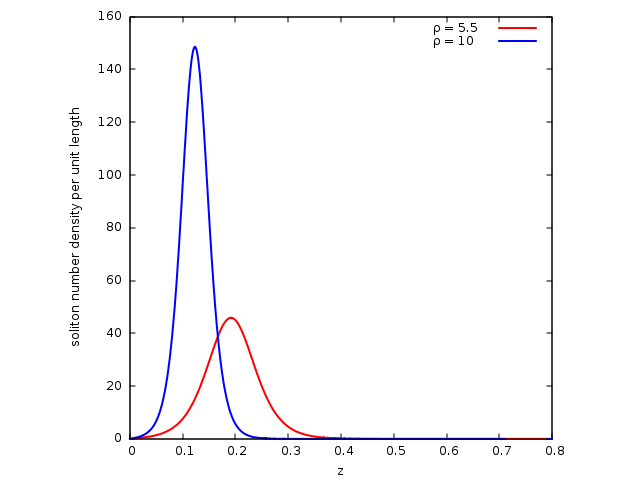}
	\caption{Plots of the numerically calculated single layer profile functions (left) and soliton number densities (right) with $\kappa=0.01$ in the homogeneous baby Skyrme model for baryon number densities $\rho=5.5$ (red) and $\rho=10$ (blue).}
	\label{2dhomogsingle}
\end{figure}

\begin{figure}[p]
	\centering
	\includegraphics[width=0.77\textwidth]{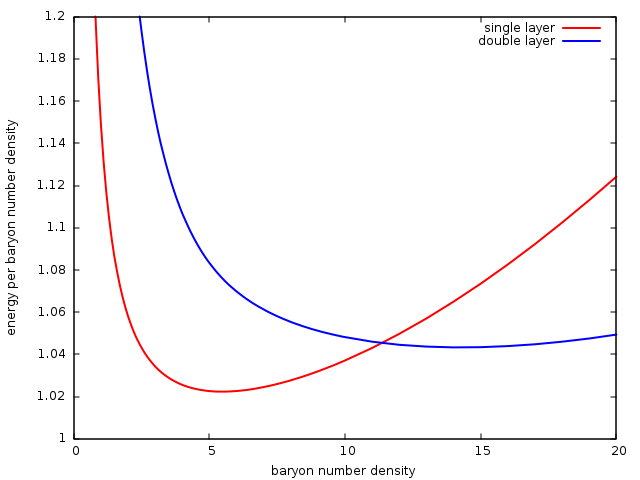}
	\caption{Energy per baryon number density (in units of $4\pi$) as a function of baryon number density $\rho$ for the single layer (red) and double layer (blue) configurations in the homogeneous baby Skyrme model with $\kappa=0.01$.}
	\label{2dhomogEnergy}
\end{figure}

It is now a straightforward exercise to numerically compute the profile functions that minimise this energy using gradient flow methods. The equation of motion for the profile function is given by
\begin{equation}\label{eqmotion2d}
\begin{aligned}
\left(1+\frac{\kappa^2\pi^2\rho^2}{H}\sin^2\!{f}\right)f^{\prime\prime} + \frac{z}{2H^2}&\left(3+\frac{\kappa^2\pi^2\rho^2}{H}\sin^2\!{f}\right)f^\prime\\ &- \frac{\pi^2\rho^2}{2H^2}\left(1-H\kappa^2(f^\prime)^2\right)\sin{2f} =0\,,
\end{aligned}
\end{equation}
although in practise it is convenient to transform the holographic direction onto the finite region $u\in[-\frac{\pi}{2}, \frac{\pi}{2}]$ via $z = \tan{u}$. Numerical solutions were obtained using a gradient flow algorithm on a grid of $1000$ points, and the corresponding profile functions and soliton number densities for $\kappa=0.01$ can be seen in Figure~\ref{2dhomogsingle}. By plotting the energy per soliton number as a function of the baryon number density $\rho$, as in Figure~\ref{2dhomogEnergy}, we also see that the optimum density for the homogeneous single layer is $\rho=5.5$. This is roughly twice the optimal density in the holographic baby Skyrme model \cite{Bolognesi:2013jba}.

We observe that the width of the homogeneous layer decreases with increasing $\rho$, which is in agreement with the full numerical results in \cite{Bolognesi:2013jba}.
This, however, seems counter-intuitive as one would expect the width of the layer to increase with increasing $\rho$. This would allow the solitonic material to explore the holographic direction, which has been proposed as a holographic description of the quarkyonic phase \cite{McLerran:2007qj} and chiral symmetry restoration \cite{deBoer:2012ij}. In fact, in the homogeneous approximation of the full Sakai-Sugimoto model \cite{Elliot-Ripley:2016uwb} it was found that the layer width does indeed increase with increasing $\rho$. This demonstrates further qualitative differences between this and the low-dimensional toy model.

To understand this effect we can look at limiting cases within the model. In the high-density, flat space limit ($\rho\to\infty$, $H=1$) the equation of motion for the profile function reduces to
\begin{equation}
\kappa^2\sin{f} f^{\prime\prime} - (1-\kappa^2(f^\prime)^2)\cos{f}=0\,,
\end{equation}
which has a compact solution
\begin{equation}\label{compactansatz}
f(z) = \begin{cases}
\pi - z/A\,, & 0\le z\le A\pi\\
0 & \text{otherwise}
\end{cases}
\end{equation}
when $A=\kappa$. Note here we have restricted to $z\ge 0$ and we extend to $z<0$ by symmetry. We can now substitute an ansatz of the form \eqref{compactansatz} into the energy \eqref{Ehomog1} and take the flat space limit to obtain
\begin{align}
E &= \int_{0}^{A\pi}\!\left\{ \frac{1}{A^2} + \pi^2\rho^2\sin^2{\left(\pi-\frac{z}{A}\right)}\left(1+\frac{\kappa^2}{A^2}\right)\right\}\,dz\,,\\
&= \frac{\pi}{2}\left( \pi^2\rho^2 A + \frac{2+\pi^2\rho^2\kappa^2}{A} \right)\,,
\end{align}
which is minimised by
\begin{equation}
A = \frac{\sqrt{2+\pi^2\rho^2\kappa^2}}{\pi\rho}\,.
\end{equation}
We note that $A\to \kappa\text{ as }\rho\to\infty$, which justifies the use of the compact ansatz. We also note that $A$ decreases as $\rho$ increases, explaining why the soliton size shrinks with increasing baryon number density.

\begin{figure}[h]
	\centering
	\includegraphics[width=0.48\textwidth]{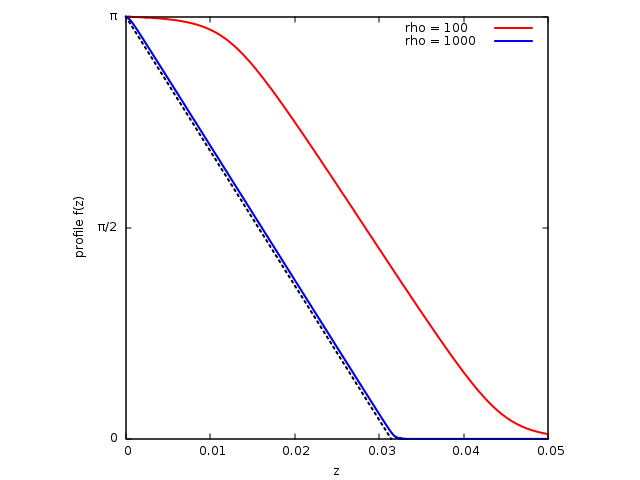}
	\includegraphics[width=0.48\textwidth]{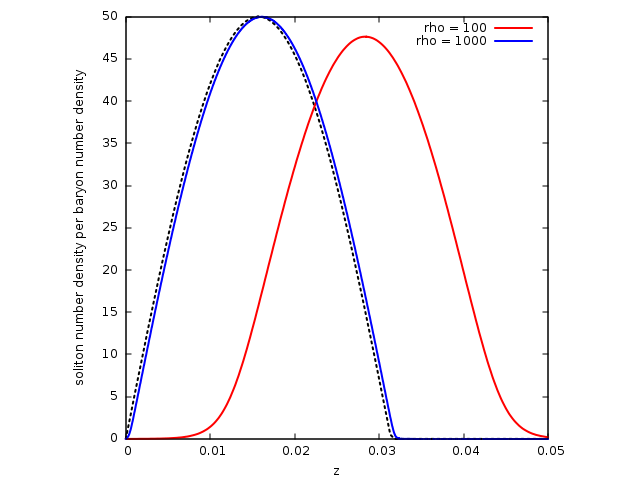}
	\caption{Plots of the numerically calculated single layer profile functions (left) and soliton number densities (normalised by the baryon number density) (right) with $\kappa=0.01$ in the homogeneous baby Skyrme model for very high baryon number densities $\rho=100$ (red) and $\rho=1000$ (blue). The black dotted lines show the associated high density compact solutions.}
	\label{vhighdensity}
\end{figure}

To determine the region of validity of the compact ansatz we must determine when the flat space approximation is applicable. The flat space approximation should be valid when the size of the soliton is small compared to the curvature scale. In other words, we require the region on which the profile function is nonzero to be in $z\in[0,\varepsilon]$, where $\varepsilon\ll 1$. In the high-density limit this yields $\kappa\pi \ll 1$, which gives
\begin{align}
\kappa \ll \frac{1}{\pi} \sim 0.3\,.
\end{align}
We see then that a value of $\kappa = 0.01$ is within this regime. Figure~\ref{vhighdensity} shows the profile function and associated soliton number density for $\rho = 100$ and $1000$, and we see that the solution does indeed approach a compact form.

\section{Baryonic popcorn in the homogeneous model}
In order to investigate the presence of baryonic popcorn in this salty homogeneous model we must first generalise our single layer ansatz to a multi layer ansatz. One straightforward way to do this for a double layer configuration, with layers localised around $z=\pm z_0$, is to patch together two single layer solutions in a continuous manner:
\begin{equation}
\pmb{\phi}(x,z) = 
\begin{cases}
\left(\sin\!{f(z)}\cos\!{(\pi\rho_1 x)}, -\sin\!{f(z)}\sin\!{(\pi\rho_1 x)}, \cos\!{f(z)}\right)\,, &\hphantom{-}z_0\le z\\
\left(\sin\!{f(z)}\cos\!{(\pi\rho_2 x)}, +\sin\!{f(z)}\sin\!{(\pi\rho_2 x)}, \cos\!{f(z)}\right)\,, &\hphantom{-}\,\,0\le z\le z_0\\
\left(-\sin\!{f(z)}\cos\!{(\pi\rho_2 x)}, +\sin\!{f(z)}\sin\!{(\pi\rho_2 x)}, \cos\!{f(z)}\right)\,, &-z_0\le z\le 0\\
\left(-\sin\!{f(z)}\cos\!{(\pi\rho_1 x)}, -\sin\!{f(z)}\sin\!{(\pi\rho_1 x)}, \cos\!{f(z)}\right)\,, &\hphantom{-z_0\le}\,\,z\le -z_0\\
\end{cases}
\end{equation}
and requiring $f(0)=f(\pm\infty)=0$ and $f(\pm z_0)=\pi$. We have introduced parameters $\rho_1$ and $\rho_2\equiv\rho-\rho_1$ to allow topological charge to distribute itself between the components of the layers, and imposed symmetry in $z\to-z$ (up to the sign flip in the second component). Again, the repeated sign flips in the second components of the field are required to ensure non-zero topological charge. There is an additional sign flip in the first and second components of the pion field (equivalent to $f(z)\to-f(z)$) between $z>0$ and $z<0$, to ensure that the two layers are out of phase with each other, to match results found in \cite{Bolognesi:2013jba}, although this has no overall effect on the computations.

We should also note that, as in the single layer case, the layers in this ansatz correspond to two layers of dyonic salt: the vanishing of the soliton number density at $z=\pm z_0$ implies that the soliton layers have already split into constituent half-layers. Again, this is a drawback compared to the analogous result in the five-dimensional Sakai-Sugimoto study, since we cannot determine whether the dyon salt or the popcorn phase is preferred at high densities. Since we have found an analogue of the baryonic popcorn phase within a dyon salt ansatz, we refer to this phase as ``salty popcorn''.

\begin{figure}[h]
	\centering
	\includegraphics[width=0.48\textwidth]{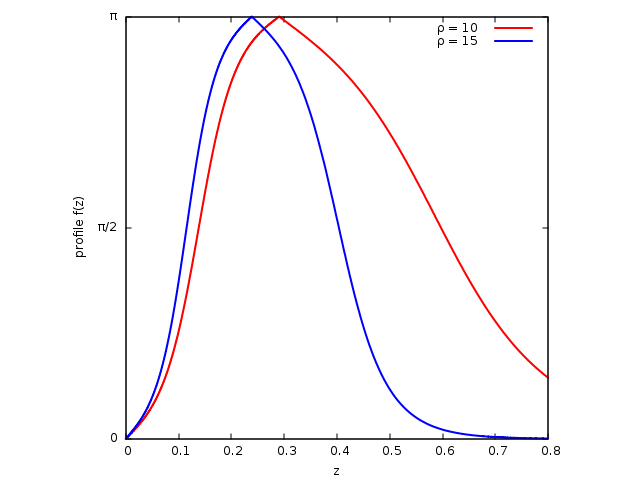}
	\includegraphics[width=0.48\textwidth]{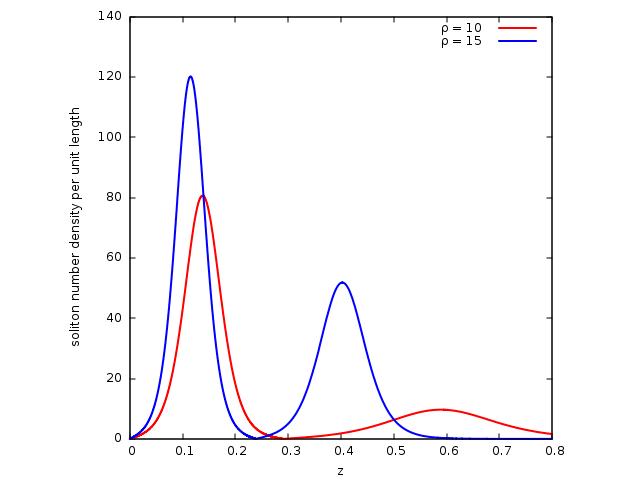}
	\caption{Plots of the numerically calculated double layer profile functions (left) and soliton number densities (right) with $\kappa=0.01$ in the homogeneous baby Skyrme model for baryon number densities $\rho=10$ (red) and $\rho=15$ (blue).}
	\label{2dhomogdouble}
\end{figure}

With this salty popcorn ansatz we can now numerically solve for the profile function and density fraction $\rho_1$ at constant $\rho = \rho_1+\rho_2$ and $z_0$ piecewise in each region using a gradient flow algorithm. We then minimise with respect to $z_0$ by performing a golden section search on these solutions. The resulting profile functions and soliton number densities are displayed in Figure~\ref{2dhomogdouble}.

Again, we observe that the widths of the layers decrease with increasing $\rho$. In addition, we find that both the separation between the layers and the relative proportion of topological charge in the inner layer both decrease with increasing baryon number density $\rho$.

We also notice that the solutions within the homogeneous ansatz become less smooth as $\rho$ increases: this could be worrying as we would expect the ansatz to become more accurate (and hence more smooth) with increasing baryon number density, as the homogeneous approximations of the five-dimensional Sakai-Sugimoto model. This can, once again, be explained by considering a compact solution in the high density, flat space limit. Restricting to $z\ge 0$ (and extending to $z<0$ using symmetry), the ansatz
\begin{equation}
f(z) = \begin{cases}
\pi z/z_0\,, & 0\le z \le z_0\\
\pi(A-z)/(A-z_0)\,, & z_0\le z\le A\\
0 & \text{otherwise}
\end{cases}
\end{equation}
again satisfies the equations of motion in the high density limit if $z_0, A-z_0\to \pi\kappa$ as $\rho_1, \rho_2\to\infty$. Substituting this ansatz into the energy \eqref{Ehomog1} and taking the flat space limit yields
\begin{equation}
\begin{aligned}
E &= \int_{0}^{z_0}\! \left\{ \frac{\pi^2}{z_0^2} + \pi^2\rho_1^2\sin^2{\left( \frac{\pi z}{z_0} \right)}\left( 1+\frac{\kappa^2\pi^2\rho_1^2}{z_0^2}\right) \right\} \,dz\\
&\qquad+ \int_{z_0}^{A}\! \left\{ \frac{\pi^2}{(A-z_0)^2} + \pi^2\rho_2^2\sin^2{\left( \frac{\pi (A-z)}{A-z_0} \right)}\left( 1+\frac{\kappa^2\pi^2\rho_2^2}{(A-z_0)^2}\right) \right\} \,dz\\
&= \frac{\pi^2}{2}\left( \rho_1^2 z_0 + \frac{2+\kappa^2\pi^2\rho_1^2}{z_0} - \rho_2^2(A-z_0) - \frac{2+\kappa^2\pi^2\rho_2^2}{A-z_0} \right)\,.
\end{aligned}
\end{equation}
Minimising with respect to $z_0$, $A$ and $\rho_1$ yields
\begin{align}
\rho_1 = \rho_2 &= \rho/2\,,\\
z_0 &= \frac{\sqrt{8+\kappa^2\pi^2\rho^2}}{\rho}\,,\\
A &= 2z_0\,.
\end{align}
We note that $z_0, (A-z_0)\to\pi\kappa$ as $\rho\to\infty$, justifying the use of the compact ansatz. We also see that we require $A = 2z_0\to 2\pi\kappa\ll 1$ for the flat space ansatz to be valid, yielding $\kappa \ll 1/(2\pi)\sim 0.16$.

\begin{figure}[h]
	\centering
	\includegraphics[width=0.48\textwidth]{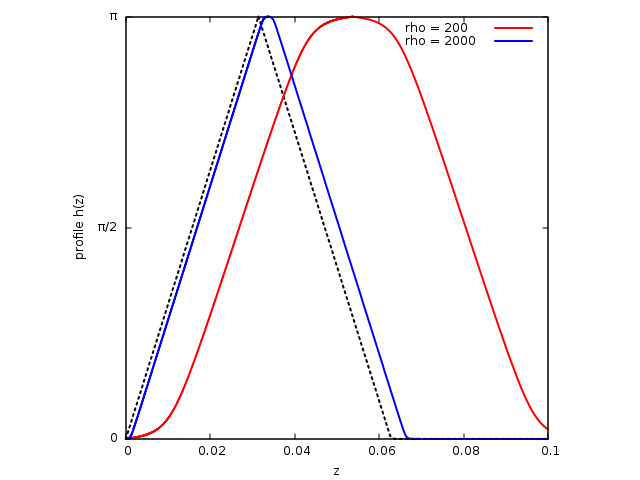}
	\includegraphics[width=0.48\textwidth]{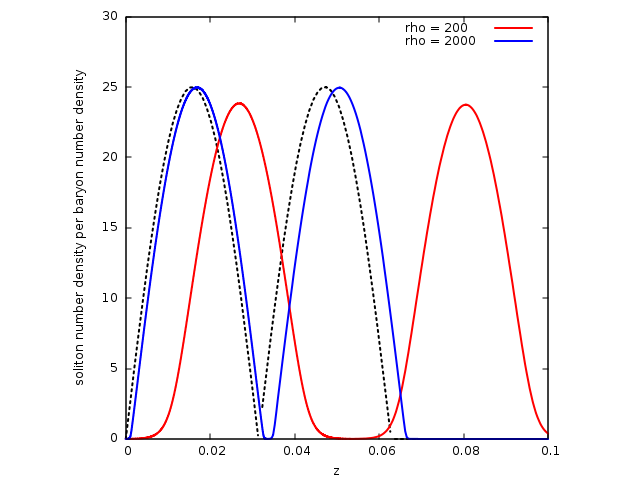}
	\caption{Plots of the numerically calculated double layer profile functions (left) and soliton number densities (normalised by the baryon number density) (right) with $\kappa=0.01$ in the homogeneous baby Skyrme model for very high baryon number densities $\rho=200$ (red) and $\rho=2000$ (blue). The black dotted lines show the associated high density compact solutions.}
	\label{vhighdensity2}
\end{figure}

We find that the high density limit of the double layer solution takes the form of two equal sized layers with baryon number density distributed equally among them. This limiting behaviour can be confirmed by numerically solving \eqref{eqmotion2d} for the double layer case for very high densities. Figure~\ref{vhighdensity2} shows the profile functions and soliton number densities for such solutions with $\kappa = 0.01$ and baryon number densities $\rho = 200$ and $\rho = 2000$. The existence of the compact solution also explains the breakdown of smoothness of the ansatz at higher densities, since the compact limit does not yield a smooth solution.

Finally, we do observe a popcorn transition in this homogeneous model. By comparing the energy per baryon number density of the single and double layer configurations in Figure~\ref{2dhomogEnergy} we can clearly see a transition point around $\rho = 11.4$ beyond which the double layer configurations become energetically favourable compared to the single layer configurations. This, again, is roughly twice the density at which the first popcorn transition is observed in the non-homogeneous baby Skyrme model. We can also see that such a popcorn transition must occur by looking at the high density compact solutions: we have
\begin{align}
E_{\text{single}} &= \pi^2\rho\sqrt{2+\pi^2\kappa^2\rho^2}\,,\\
E_{\text{double}} &= \frac{\pi^2\rho}{2}\sqrt{8+\pi^2\kappa^2\rho^2}\,,
\end{align}
and so we see that $E_{\text{double}}\to\frac{1}{2}E_{\text{single}}$ as $\rho\to\infty$.

\section{Comparison with full numerical solutions}
The advantage of the two-dimensional toy model is that we can actually compare the predictions made by the homogeneous ansatz to numerical solutions obtained in the full model.
In \cite{Bolognesi:2013jba} full numerical periodic solutions in the holographic baby Skyrme model were found for a range of baryon densities, with boundary conditions used to specify different numbers of layers. An analogue of the dyon salt phase was found, manifest by single layer solutions becoming more homogeneous with increasing density. In addition, double and triple layer configurations were found to be energetically favourable beyond some critical densities, demonstrating popcorn transitions. Crucially, these popcorn transitions occurred at densities well below the analogues of the dyon salt. It was conjectured that further popcorn transitions beyond three layers would also occur.

\begin{figure}[p]
	\centering
	\includegraphics[width=0.48\textwidth]{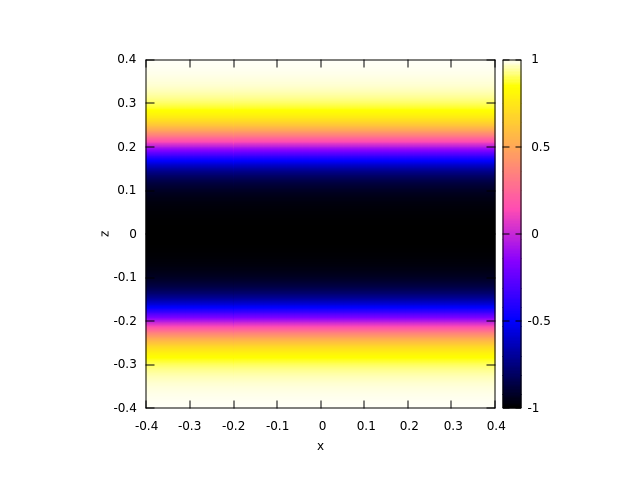}
	\includegraphics[width=0.48\textwidth]{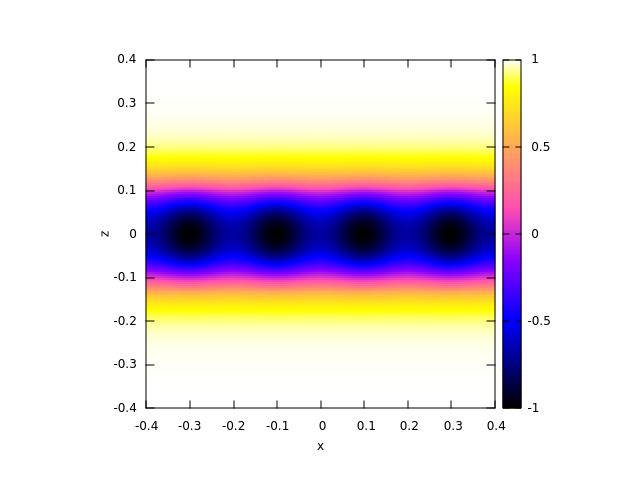}
	\hfill
	\includegraphics[width=0.48\textwidth]{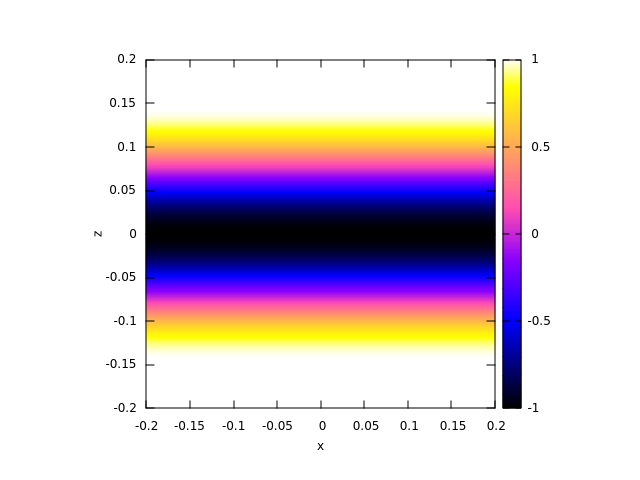}
	\includegraphics[width=0.48\textwidth]{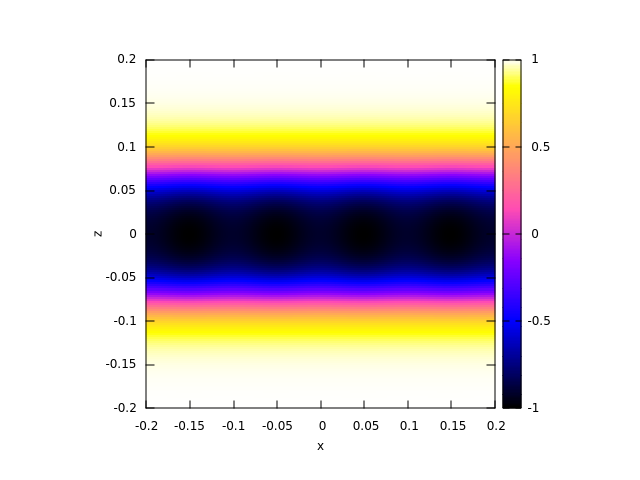}
	\caption{The third component of the pion field $\phi_3$ for the single layer configurations within the homogeneous ansatz (left) and from full numerical computations (right). Each plot shows four units of topological charge. The solutions are calculated at densities $\rho=5$ (top) and $\rho=10$ (bottom). We see that the homogeneous approximation is reasonable for moderately high densities.}
	\label{superpops1}
\end{figure}

\begin{figure}[p]
	\centering
	\includegraphics[width=0.48\textwidth]{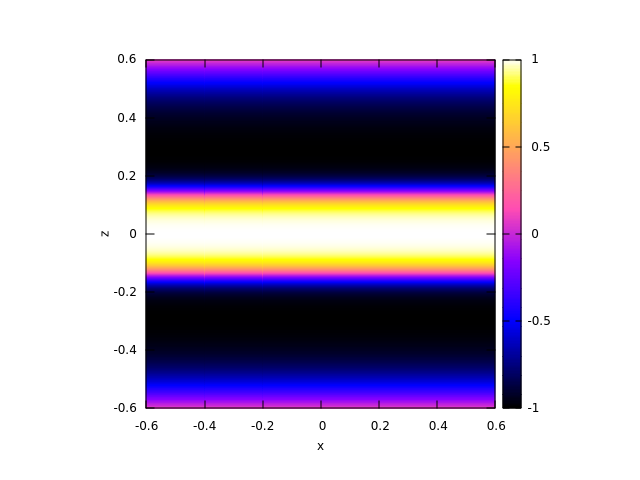}
	\includegraphics[width=0.48\textwidth]{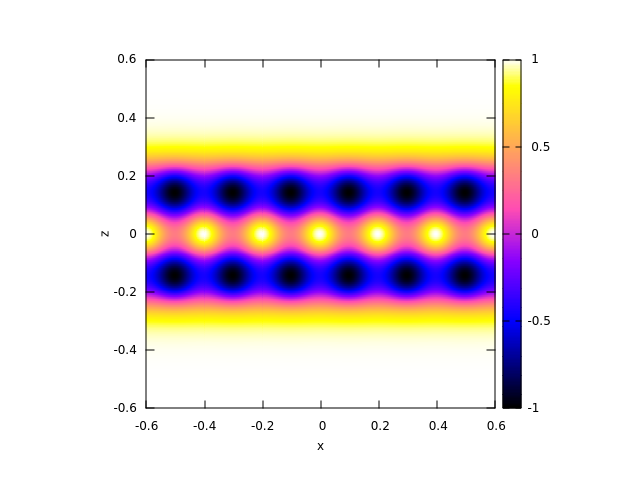}
	\hfill
	\includegraphics[width=0.48\textwidth]{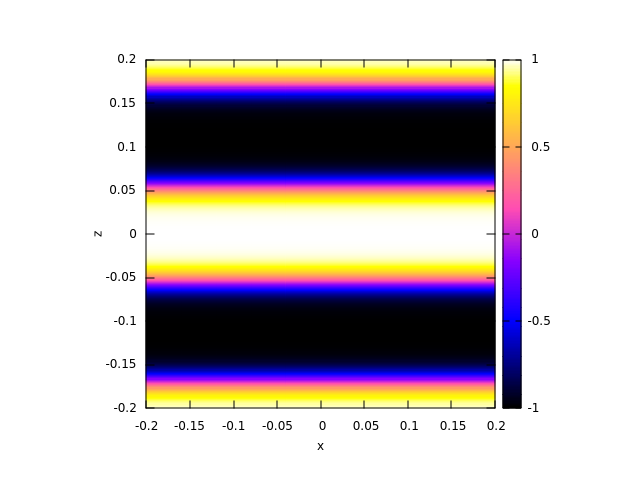}
	\includegraphics[width=0.48\textwidth]{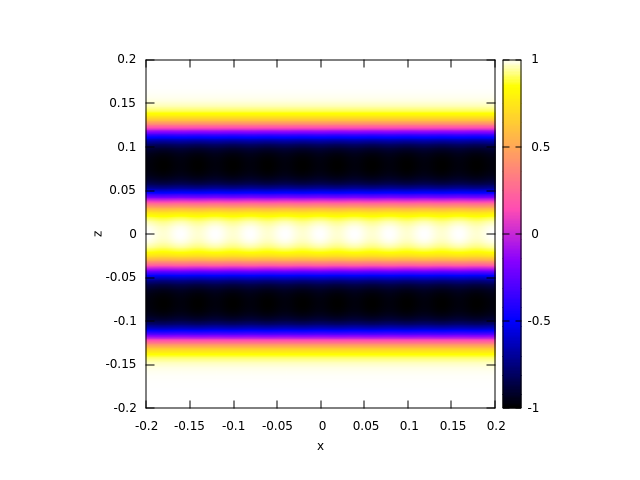}
	\caption{The third component of the pion field $\phi_3$ for the double layer configurations within the homogeneous ansatz (left) and from full numerical computations (right). The solutions are calculated at densities $\rho=10$ (top, $12$ units of topological charge displayed) and $\rho=50$ (bottom, $20$ units of topological charge displayed). We see that the homogeneous approximation is reasonable for moderately high densities.}
	\label{superpops2}
\end{figure}

Figure~\ref{superpops1} shows the pion field $\phi_3$ for both the homogeneous ansatz and the full numerical solution for single layer configurations at densities $\rho=5$ and $\rho=10$, and Figure~\ref{superpops2} shows the same information for double layer configurations at densities $\rho=10$ and $\rho=50$. In each case, numerical computations were performed on a grid containing one unit of topological charge per layer and periodic boundary conditions on $\phi_3$. Anti-periodic boundary conditions on $\phi_1$ and $\phi_2$ were imposed to ensure solutions took the form of either a single chain or a regular square chain respectively (as in the multi-layer computations in \cite{Elliot-Ripley:2015cma}). The plots themselves show more units of topological charge, for the sake of clarity.

In both cases we find the qualitative appearances of the solutions to be very similar for higher densities. We also calculate the energies of the different solutions (displayed in Table~\ref{superpopenergy}) and find them to also be in close agreement at higher densities. This not only lends credibility to the homogeneous ansatz as used to investigate the toy model, but also gives some evidence that the homogeneous approximation used in \cite{Elliot-Ripley:2016uwb} may be able to qualitatively capture the behaviour of the five-dimensional Sakai-Sugimoto model at high densities.

\begin{table}[p]
	\centering
	%
	\begin{tabular}{ | c | c | c | c | c | }
		\hline
		\multirow{2}{*}{density $\rho$} & \multicolumn{2}{c|}{single layer $E/(4\pi\rho)$} & \multicolumn{2}{c|}{double layer $E/(4\pi\rho)$} \\
		\hhline{~----}
		{} & homogeneous & numerical & homogeneous & numerical \\
		\hline                       
		5 & 1.023 & 1.012 & 1.088 & 1.013 \\
		10 & 1.037 & 1.034 & 1.048 & 1.021 \\
		50 & 1.608 & 1.608 & 1.189 & 1.187 \\
		
		\hline
	\end{tabular}
	\caption{Energies per baryon number density (in units of $4\pi$) for single and double layer solutions, both within the homogeneous ansatz and from full numerical computations.}
	\label{superpopenergy}
\end{table}

We also notice that, for the double layer configurations, the homogeneous ansatz predicts the layers to be more widely separated, even for higher densities. This could be because the homogeneous ansatz does not allow the separate layers to interact, as the ansatz forces the fields to take the vacuum value on the line $z=0$. In the full numerical solutions the separate layers are out of phase with each other, causing an attraction between them, and the homogeneous ansatz cannot capture this feature.

\section{Multi-layer solutions and popcorn transitions}
We can use our flat space compact approximations to shows the existence of further popcorn transitions in this model beyond two layers. For simplicity we will consider an $N$-layer solution for $N$ odd, although this can be easily generalised for $N$ even. Restricting to $z\ge 0$ we construct a multi-layer ansatz with soliton layers localised around the positions $z\in\{ z_0=0, z_1, \dots, z_{(N-1)/2}\}$ by piecewise gluing together compact solutions of the form above. Explicitly we write
\begin{equation}
f(z)=\begin{cases}
\pi(A_n-z)/(A_n - z_n)\,, & z_n\le z \le A_n\\
\pi(z-A_n)/(z_{n+1}-A_n)\,, & A_n\le z \le z_{n+1}\\
0 & \text{otherwise}
\end{cases}\,,
\end{equation}
where $0\le n \le (N-1)/2$ and $z_n<A_n<z_{n+1}$. This ensures $f(z_n) = \pi$ and $f(A_n)=0$. This ansatz can then be substituted into the energy \eqref{Ehomog1}, which can be integrated in the flat space limit to obtain
\begin{equation}
E = \sum_{k=0}^{N-1} E_{k}\,,
\end{equation}
where we denote the partial energies as
\begin{align}
E_{2k} &= \frac{\pi^2}{2}\left( (A_k-z_k)\rho_{2k}^2 + \frac{2+\kappa^2\pi^2\rho_{2k}^2}{A_k-z_k}\right)\,, &0\le k\le (N-1)/2\\
E_{2k+1} &= \frac{\pi^2}{2}\left( (z_{k+1}-A_k)\rho_{2k+1}^2 + \frac{2+\kappa^2\pi^2\rho_{2k+1}^2}{z_{k+1}-A_k}\right)\,, &0\le k\le (N-3)/2
\end{align}
and $\rho_k$ are the partial baryon number densities satisfying $\rho = \sum_{k=0}^{N-1}\rho_k$.

We can now minimise with respect to $z_n$ and $A_n$ to obtain
\begin{align}
A_k-z_k &= \frac{\sqrt{2+\kappa^2\pi^2\rho_{2k}^2}}{\rho_{2k}}\,,& &0\le k\le (N-1)/2\\
z_{k+1}-A_k &= \frac{\sqrt{2+\kappa^2\pi^2\rho_{2k+1}^2}}{\rho_{2k+1}}\,,& &0\le k\le (N-3)/2\\
E_k &= \pi^2\rho_k\sqrt{2+\kappa^2\pi^2\rho_k^2}\,,& &0\le k \le N-1
\end{align}
Further minimising with respect to $\rho_k$ yields $\rho_k = \rho/N$ and we find that, within this ansatz, the baryon number density of each layer component is equal. This then gives us
\begin{align}
A_k-z_k = z_{k+1}-A_k &= \frac{\sqrt{2N^2+\kappa^2\pi^2\rho^2}}{\rho}\,,\\
E = \sum_{k=0}^{N-1}E_k &= \frac{\pi^2\rho}{N}\sqrt{2N^2+\kappa^2\pi^2\rho^2}\,,
\end{align}
and we see that the widths of each layer component are also all equal, tending to $\kappa\pi$ as $\rho\to\infty$. As above, this implies that this ansatz solves the flat space equation of motion in the high density limit. The flat space approximation is valid when $N\kappa\pi\ll 1$, which for our value of $\kappa=0.01$ gives $N\ll 1/(0.01\pi) \sim 32$. We can also see that, as long as the flat space approximation is valid, the $N$-layer solution will always be preferable in the high density limit: we have $E_{\text{N-layer}} \to \frac{1}{N}E_{\text{single}}$ as $\rho\to\infty$.

The analysis above is only valid when the flat space approximation is legitimate; beyond this regime the effect of curved space would need to be taken into account. This is difficult both numerically and analytically due to the large number of extra parameters we have introduced. We can, however, construct an ansatz for three layers subject to further approximations, and deduce the existence of further popcorn transitions.

\begin{figure}[t]
	\centering
	\includegraphics[width=0.48\textwidth]{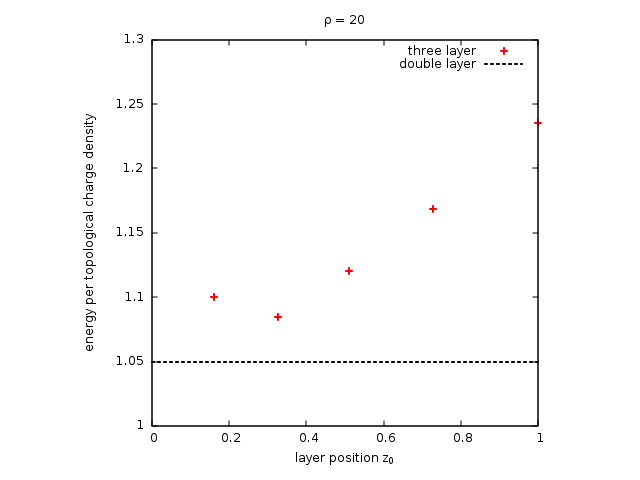}
	\includegraphics[width=0.48\textwidth]{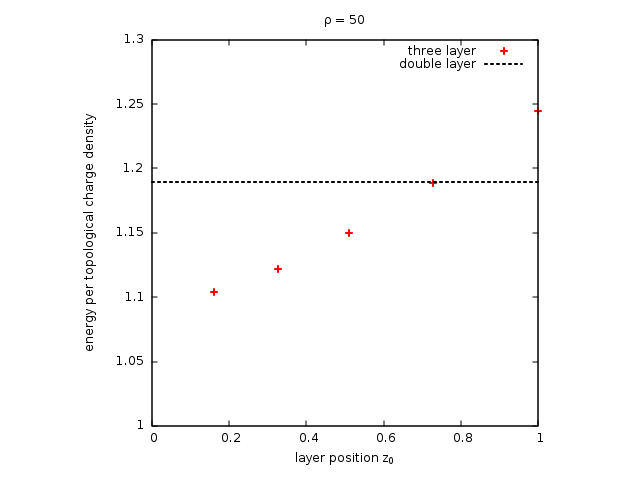}
	\caption{Energy per baryon number density (in units of $4\pi$) as a function of layer position $z_0$ for $\rho=20$ (left) and $\rho=50$ (right) in the homogeneous baby Skyrme model with $\kappa=0.01$. The black dotted lines show the corresponding double layer energies at these densities, and the energies of the three layer configurations are given by red $+$'s.}
	\label{threelayer}
\end{figure}

Using simpler notation than the multi layer analysis above, we construct a three layer ansatz by introducing parameters $z_1>z_0$ and imposing $f(0)=f(\pm z_1)=\pi$ and $f(\pm z_0)=f(\pm\infty)=0$ (with appropriate sign flips in the second components of the associated $\pmb{\phi}$ fields). Numerically minimising with respect to all of these parameters, in addition to the partial baryon number densities in each layer component, is challenging. Instead, we assume the baryon number density is equally spread across the configuration (by setting all the partial baryon number densities equal to $\rho/3$) and then minimise with respect to $z_1$ for a range of fixed $z_0$. By doing this we can clearly see that for high densities, there will be a three layer solution that is energetically preferable when compared to the double layer (see Figure~\ref{threelayer}).

\section{Conclusions}
We have successfully employed a homogeneous ansatz which has reproduced some of the qualitative behaviour of the full holographic baby Skyrme model. We find a series of single layer configurations, with an optimal density and a popcorn transition (albeit a salty one) to a double layer configuration at roughly twice the true numerically calculated densities. We also find evidence for a second popcorn transition in this model, qualitatively matching the full numerical results previously studied.
These successes demonstrate that a homogeneous ansatz can be a useful tool for studying cold, dense QCD. This also serves as evidence to suggest that the true high-density solutions in the full Sakai-Sugimoto model may take the form of the soliton bag found in \cite{Elliot-Ripley:2016uwb}, rather than the previously proposed dyon salt or multiple layers of baryonic popcorn.

Unfortunately the ansatz falls short on multiple counts. Firstly, the ansatz can only generate solutions that are already in the dyon salt phase, preventing any comparisons between the energy of this phase with the energy of the baryonic popcorn phase.
In fact, we have also shown that imposing any homogeneity in this toy model via periodic fields will always have this problem. A low-dimensional analogue of the kink ansatz of \cite{Elliot-Ripley:2016uwb} would be a more useful approximation, but this does not seem possible in the low-dimensional case.


As a final comment, we have found that the low-dimensional toy models for the Sakai-Sugimoto model behave qualitatively differently from the five-dimensional model. There are a number of reasons why this may occur. Firstly, the toy model only has two spatial dimensions, as opposed to the four spatial dimensions of Sakai-Sugimoto, which could play a role. Secondly, the toy model uses a sigma model to approximate a gauge theory, and the differences between these classes of model could have significant effects. Finally, looking back to the vector meson model studied in \cite{Elliot-Ripley:2015cma} we see that the vector meson field, which plays the role of the abelian electric potential $\hat{A}_0$ in the Sakai-Sugimoto model, must have a nonzero mass term to allow for finite energy configurations, whereas there is no corresponding mass term in the Sakai-Sugimoto model. The presence of this mass means that the homogeneous equation of motion for the vector meson is no longer once-integrable, and the field cannot be eliminated from the Lagrangian as it was in \cite{Elliot-Ripley:2016uwb}. 

\section{Acknowledgements}
This work is supported by an STFC PhD studentship. I would also like to thank Marija Zamaklar and my supervisor, Paul Sutcliffe for their support and useful discussions.

\appendix
\section{Appendix: Salt as a result of homogeneity and periodicity in the baby Skyrme model}\label{ap:salt}
Here we show that a salty ansatz is unavoidable when imposing periodicity in the (holographic) baby Skyrme model with the boundary conditions $\pmb{\phi}(x,0)=(0,0,-1)$ and $\pmb{\phi}(x,\infty)=(0,0,1)$. Note that the boundary condition at $z=0$ corresponds to requiring a layer of solitonic material located there, and is unique to making the homogeneous ansatz. It is not a general feature of periodic baby Skyrmions.

The general form for the pion field $\pmb{\phi}$ is given by
\begin{equation}
\begin{aligned}
	\pmb{\phi}(x,z) = \left(f_1(x,z), f_2(x,z), \pm\sqrt{1-f_1^2-f_2^2}\right)\,,
\end{aligned}
\end{equation}
and periodicity can be imposed in the $x$ direction by expanding the functions $f_1$ and $f_2$ as Fourier series:
\begin{equation}
\begin{aligned}
f_1(x,z) &= \sum_{n\in\mathbb{Z}} \alpha_n(z)e^{2in\pi x/L}\,, \\
f_2(x,z) &= \sum_{m\in\mathbb{Z}} \beta_m(z)e^{2im\pi x/L}\,.
\end{aligned}
\end{equation}
The boundary conditions then imply $\alpha_n(0)=\alpha_n(\infty) = \beta_m(0)=\beta_m(\infty) =0$. The topological charge density of such a configuration is then given by
\begin{align}
\mathcal{B} &= -\frac{1}{4\pi}\pmb{\phi}\cdot\left( \partial_x\pmb{\phi}\times\partial_z\pmb{\phi} \right)\\
&= \frac{-1}{4\pi\sqrt{1-f_1^2-f_2^2}}\left(\partial_xf_1\partial_zf_2 - \partial_xf_2\partial_zf_1\right)\\
&= \frac{-1}{2\sqrt{1-f_1^2-f_2^2}}\sum_{n,m}\frac{in}{L}\left(\alpha_n\beta_m^\prime - \alpha^\prime_m\beta_n\right)e^{2i(n+m)\pi x/L}\,.
\end{align}
We then see that the boundary conditions imply that $\mathcal{B}$ must vanish at $z=0$ and $z=\infty$. In other words, the topological charge density, and therefore baryon number density, must be localised around a point away from $z=0$. This means that, by imposing periodicity in the nonholographic direction, we have forced the homogeneous ansatz to be salty.

\bibliography{../../Papers/master.bib}
\bibliographystyle{../../Papers/JHEP}

\end{document}